# Wetting on smooth micropatterned defects

Damien Debuisson, Renaud Dufour, Vincent Senez and Steve Arscott*

*Institut d'Electronique, de Microélectronique et de Nanotechnologie (IEMN),
University of Lille, CNRS UMR-8520, Cite Scientifique, Villeneuve d'Ascq, France.*

We develop a model which predicts the contact angle hysteresis introduced by smooth micropatterned defects. The defects are modeled by a smooth function and the contact angle hysteresis is explained using a tangent line solution. When the liquid micro-meniscus touches both sides of the defect simultaneously, depinning of the contact line occurs. The defects are fabricated using a photoresist and experimental results confirm the model. An important point is that the model is *scale-independent*, i.e. the contact angle hysteresis is dependent on the aspect ratio of the function, not on its absolute size; this could have implications for natural surface defects.

PACS numbers: 68.08.Bc, 47.55.D-, 85.40.Hp

Complex wetting [1-5] is observed on both natural [6-8] and artificial [2,5,9] surfaces. When a sessile droplet contracts (e.g. via evaporation) or swells (e.g. via condensation) the contact line (i.e. the liquid-vapor-solid interface) can move. The contact angle hysteresis [3,10-13] is defined as the difference between the advancing contact angle (i.e. the contact line moving forwards) and the receding contact angle (i.e. the contact line moving backwards) and remains a poorly understood phenomenon. Here we show that the contact angle hysteresis introduced by a "smooth, regular" [4,12] defect can be successfully predicted using a simple *scale-independent* geometrical model. Contact angle hysteresis tuning could be fundamental for modern applications such as *inter alia* self-cleaning surfaces [14], miniaturized chemistry [15], self-assembly [16] and nanotechnology [17].

We consider here the contact angle hysteresis introduced by a smooth defect [4,12]. We assume that the defect can be modeled by a function $h(x)$ which is smooth in the mathematical sense i.e. it is differentiable. Fig. 1 illustrates the four possible cases considered here.

Figure 1(a) is an *advancing* contact line moving from left to right which encounters a smooth defect where the micro-meniscus *does not touch* the right hand side (r.h.s.) of the defect as the contact line moves over the defect. Fig. 1(b) is an *advancing* contact line, again moving from left to right, which encounters a smooth defect where the micro-meniscus *does touch* the r.h.s. of the defect. Fig. 1(c) is a *receding* contact line moving from right to left which encounters a smooth defect where the micro-meniscus *does not touch* the left hand side (l.h.s.) of the defect. Fig. 1(d) is a *receding* contact line moving from right to left which encounters a smooth defect where the micro-meniscus *does touch* the l.h.s. of the defect. In all cases the centre of the droplet is considered to be to the left of the defect. We consider that far from the defect the *moving* contact line has a receding or an advancing contact angle [4,12] given by $\theta_{r,a}$ and we define the *effective* receding and advancing contact angles $\theta_{r,a}^*$ as the contact *angle when the contact line depins from the defect*. Let us now consider the case of the receding contact angle in more detail, although it will be seen that the arguments apply equally to the advancing contact angle via the symmetry of the defect considered here. As the contact line moves from right to left far from the defect, it does so with a contact angle equal to the receding contact angle $\theta_r$. As the contact line moves down the r.h.s. ($x > 0$) of the defect, $\theta_r$ is maintained locally (in keeping with Young-Dupré equation [18,19]) but as we have seen the macroscopic contact angle reduces. It is important now to make a physical distinction between the liquid micro-meniscus touching the l.h.s. ($x < 0$) of the defect or not.

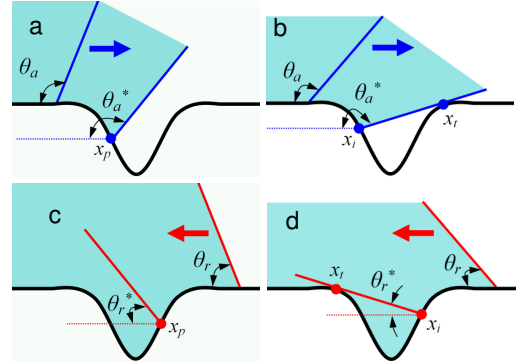

FIG. 1. Illustration of a contact line moving across a smooth defect. (a) The micro-meniscus of an advancing contact line (blue) does not touch the r.h.s. of the defect. (b) The micro-meniscus of an advancing contact line (blue) does touch the r.h.s. of the defect. (c) The micro-meniscus of a receding contact line (red) does not touch the l.h.s. of the defect. (d) The micro-meniscus of a receding contact line (red) does touch the l.h.s. of the defect.

If $\theta_r$ is large and/or the defect is shallow such that the micro-meniscus does not touch the opposite side of the defect (Fig 1(c)) then the *effective* receding contact angle *introduced by the defect* is given by $\theta_r^* = \theta_r - \tan^{-1}(h'(x_p))$ where $x_p$ is the point of inflection on the r.h.s. of $h(x)$; the relationship between $\theta_r$ and $\theta_r^*$ is linear. Movement of the contact line past this point would result in instability as the local curvature of the liquid at the defect would be higher than the curvature of the droplet; this situation conflicts with the Young-Laplace equation [19,20] and would result in the microscopic contact angle being greater than the macroscopic contact angle. The contact angle hysteresis *of the defect* $\Delta\theta^*$ is thus given by $\Delta\theta + 2\tan^{-1}(h'(x_p))$ where $\Delta\theta = \theta_a - \theta_r$; this is the contact angle hysteresis far from the defect.

If, however, as the contact line moves down the r.h.s. of the defect with the macroscopic contact angle reducing, the liquid meniscus *touches* the l.h.s. of the defect (at $x = x_t$) then in this case $\theta_r^*$ of the defect, which is given by $\tan^{-1}(h'(x_t))$, is equal to $\theta_r - \tan^{-1}(h'(x_i))$; the relationship between $\theta_r$ and $\theta_r^*$ is non-linear. This is the *tangent line* at $x_t$ which also intersects $h(x)$ at $x = x_i$; this case was illustrated in Fig. 1(d). By symmetry, $\theta_a^*$ is given by $\theta_a - \tan^{-1}(h'(x_i))$ by using the same arguments above but with the contact line moving from left to right (Fig. 1(b)). Continuing the receding angle case, once the micro-meniscus touches the l.h.s. of the defect at $x_t$ it will break as the Young-Dupré equation [18,19] is not satisfied at $x_t$; as a consequence the droplet will depin. As $h(x)$ is a smooth function, it follows that we should be able to predict the contact angle hysteresis introduced by the defect in terms of simple parameters (the defect depth $h_0$, the defect width $w$ and $\theta_{r,a}$). As a first approximation, since $w$ and $h_0$ are much smaller than the droplet diameter (~1 mm) we can consider the profile of the micro-meniscus at the defect to be that of a straight line (although in reality it is a curve). Thus, the required tangent line cuts $h(x)$ at two points: the tangent point $(x_t)$ on the l.h.s. of the defect and the intersection point $(x_i)$ on the r.h.s. of the defect. Hence, we can write $\theta_{r,a}^* = \tan^{-1}(h'(x_t))$ and $\tan^{-1}(h'(x_t)) = \theta_{r,a} - \tan^{-1}(h'(x_i))$ together with the tangent line $h(x_t) - h(x_i) = h'(x_t)(x_t - x_i)$.

The micropatterned defects can be modeled using a cosine function over a single period: $h(x) = -\frac{1}{2}(\cos(2\pi x/w)+1)$ where $h_0$ and $w$ are described above. Two tangent line solutions exist for given values of $\theta_r$ and $\theta_a$; the required solution is the tangent line which interests $h(x)$ *above* its point of inflection as this is where the micro-meniscus initially touches both sides of the defect simultaneously. A numerical solution to the equations above leads to the curves given in Fig. 2 which shows the calculated *effective* receding and advancing contact angles as a function of the receding and advancing contact angles.

Due to symmetry, the tangent line solution applies to both receding and advancing contact angles as $\theta_a^* = \pi - \theta_r^*$; note that this only holds for a defect which is symmetrical around $x = 0$. Note that Fig. 2 indicates that there can be two values of $\theta_{r,a}^*$ for a given $\theta_{r,a}$; the dashed lines correspond to the second tangent line solution and do not physically occur as the micro-meniscus touches the other side of the defect and depins at the first tangent line solution. It should also be noted that the solutions are *scale-independent*, i.e. a tangent line solution is true for *any* cosine function (defect) having a ratio $h_0/w = \kappa$ where $\kappa$ is the *aspect ratio* of the defect, as shown in the inset of Fig. 2.

In order to test our model we fabricated micropatterned surfaces using the photoresist SU-8. A photolithographic mask enabled the formation concentric circles (trench-like defects) of diminishing diameters (0.1-2 mm), see Fig. 3. The surfaces were also covered with a thin coating (~100 nm) of fluorocarbon (FC) [21] using a $C_4F_8$ plasma deposition (Surface Technology Systems, UK).

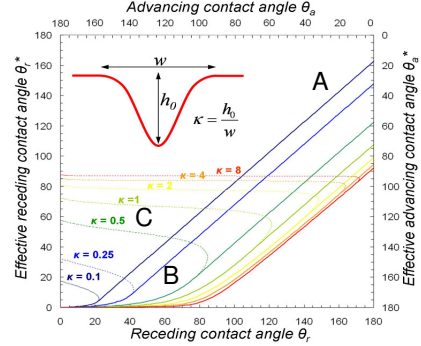

FIG. 2. Theoretical curves for the effective receding and advancing contact angles on defects of differing aspect ratio $\kappa$. The linear part (indicated by A) of the curve (slope = 1) corresponds to the case where the micro-meniscus does not touch the opposite side of the defect when moving across it. The non-linear part of the curve (indicated by B) corresponds to the first tangent line solution. The dashed non-linear part of the curve (indicated by C) corresponds to the second tangent line solution which is not physically observed here. Inset shows the dimensions of the defect: $h_0$ corresponds to the defect depth, $w$ the width and $\kappa$ the aspect ratio.

The measurements were performed in a class ISO 5/7 cleanroom ($T_A$ = 20°C±0.5°C; $RH$ = 45%±2%) using a contact angle meter (Kruss, Germany). Contact angle hysteresis was measured using droplet expansion (syringe pump) and evaporation [22-24]. In order to measure $\theta_a^*$ and $\theta_r^*$ using the syringe method, a small droplet of liquid (*vol* ~0.3 nL) was positioned onto the smallest concentric circle ($\varphi = 100$ μm) at the centre of the micropatterned surface and swelled using a syringe pump (1-5 μL min$^{-1}$). Once the droplet diameter was larger than the largest concentric circle in the micropattern ($\varphi = 2$ mm), the droplet volume was reduced using the syringe (1-5 μL min$^{-1}$) in order to measure $\theta_r^*$. For the evaporation tests, a droplet (*vol* ~2 μL) was placed on the micropatterned surface and allowed to slowly evaporate (unforced) whilst the value of $\theta_r^*$ was recorded as a function of time. This evaporation method was not possible for ethylene glycol due to a low vapor pressure at room temperature [25]; however, evaporation tests were possible using IPA/H$_2$O solutions as the vapor pressure of IPA and of water are very similar at room temperature [25].

Fig. 3(a) shows a typical droplet evaporation sequence recorded for a micropattern. The droplet evaporates until it becomes pinned onto a circular defect (t = 431 s). The apparent contact angle reduces until the effective contact angle is attained whereby the droplet suddenly depins (t = 661 s) and pins to the subsequent circle defect, the process repeats one more time. Fig. 3(b) shows data gathered over a period of 375 s for an evaporating droplet (IPA/H$_2$O: 2.5/97.5) on a FC coated SU-8 surface containing smooth defect concentric circles spaced by 100 µm. Two distinct phases are apparent: Phase (i) corresponds to droplet evaporation at constant contact angle with diminishing droplet base radius [23] when the droplet is not pinned to a defect and phase (ii) involves an evaporating droplet depinning from circle defect to circle defect. The droplet depinning is described by the equation in Fig. 3(b). However, between depinnings, which occurs very quickly (<200 ms) at constant volume (cf. Fig. 3(a)) the evaporation is described as in Ref. [23]. The model presented here (see Fig. 2) predicts the depinning contact angle $\theta_r^*$; we now test this model with various liquids.

Fig. 4 shows a plot of the experimentally obtained values of $\theta_r^*$ and $\theta_a^*$ as a function of $\theta_r$ and $\theta_a$ for various liquids on surfaces containing smooth micropatterned defects. Our measurements indicate that the model can accurately predict the values of $\theta_r^*$ introduced by the defect thus giving strong evidence for the tangent line solution model for the contact angle hysteresis introduced by a smooth defect. Fig. 4(a) indicates that there are six points lying on the non-linear solution and one point (red circle) on the linear solution. The inset to Fig. 4(a) illustrates good agreement between the experimental defect and the model. Despite the defect not being a perfect cosine shape the parameters $h_0$ and $w$ enable a good fit between the real defect shape and the model *in the region where the liquids depin*.

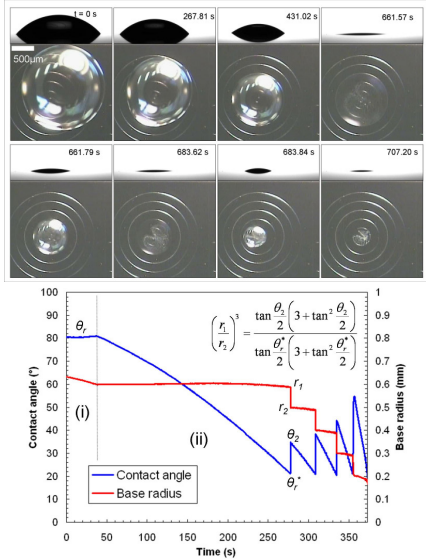

FIG. 3. (a) Plan and side view optical images of a liquid droplet (H$_2$0) evaporation sequence taken over 707 s on a micropatterned surface (SU-8) composed of concentric circle defects separated by 200 µm. (b) Variation of contact angle (red line) and droplet base diameter (blue line) as a function of time for a IPA/H$_2$O (*vol/vol* = 2.5/97.5) droplet on smooth defects separated by 100 µm defined in SU-8 covered with FC.

Let us now consider the experimental results for $\theta_a^*$ shown in Fig. 4(b). Due to the finite needle diameter the values of $\theta_a^*$ were measured at droplet base radii between 0.5-1 mm. Two liquid/surface combinations fit very well with the predictions of the model and five deviate from the model. In order to explain this, horizontal colored bars have been added to Fig. 4(b) which indicate the value of contact angle calculated using a base radius $r$ of 0.75 mm and equating the droplet diameter $D$ to the capillary length $(\gamma/\rho g)^{1/2}$ of the liquid. These bars represent the point where gravity can no longer be considered to be negligible; droplets can vibrate [26] leading to possible depinning [27] before the tangent line solution. To illustrate this another way, in order to observe $\theta_a^* = 175.3°$, as predicted by the model, then a droplet of ethylene glycol (on a FC covered defect, $r = 0.75$ mm, $\gamma = 47.7$ mJ m$^{-2}$, $\rho = 1096.8$ kg m$^{-3}$ [25]) would need to be swelled to a diameter of 18.3 mm.

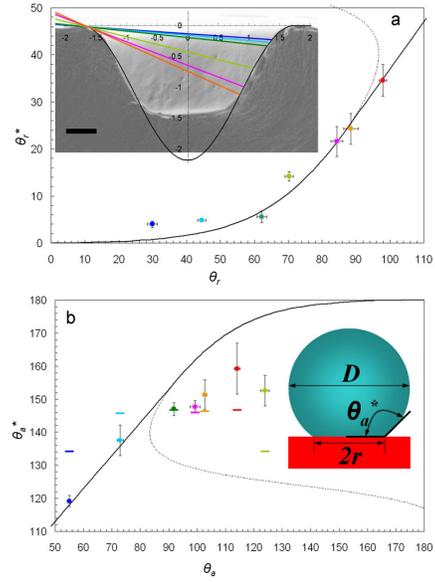

FIG. 4. Experimental values of the effective receding and advancing contact angles on a defect for various liquids. (a) The effective receding contact angle ($\theta_r^*$). H$_2$0 on SU-8 (dark green circle), H$_2$0 on FC (red circle), C$_2$H$_6$O$_2$ on SU-8 (dark blue circle), C$_2$H$_6$O$_2$ on FC (light green circle), IPA/H$_2$O (2.5/97.5) on FC (orange circle), IPA/H$_2$O (5/95) on FC (pink circle) and IPA/H$_2$O (10/90) on SU-8 (light blue circle). The predicted curve (black) using the model is also shown (using an aspect ratio $\kappa = 0.625$). Inset shows an SEM image of a cross-section of a defect formed using photolithography of SU-8 (the 100 nm thick fluorocarbon layer deposited onto the SU-8 surface is visible) superimposed with the model cosine curve with $\kappa = 0.625$. The colored lines represent the tangent line solutions for the various liquids; the color of the lines corresponds to the code above. (b) The effective advancing contact angle ($\theta_a^*$) on a defect using the same color code as above. The predicted curve (black) using the model is also shown (using an aspect ratio $\kappa = 0.625$). The colored bars (same color code as above) represent the calculated contact angle when the droplet diameter is equal to the capillary length of the liquid. Inset show a droplet resting on a solid surface, $D = 2r/\sin(\pi-\theta_a^*)$.

As a summary, the table shows the receding contact angle $\theta_r$ and the advancing contact angle $\theta_a$ on SU-8 and C$_4$F$_8$ treated

SU-8 with and without the presence of smooth defects using deionized water (DI-H$_2$O), isopropyl alcohol (IPA) [VLSI grade, Carlo Erba Reagents, Italy]/DI-H$_2$O solutions (*vol/vol* = 2.5/97.5, 5/95 and 10/90) and ethylene glycol (C$_2$H$_6$O$_2$) [Reagent grade, Scharlau, Spain]. The experimental values were obtained using tilt and syringe (contraction and expansion) and methods. All measurements were conducted at least 5 times.

Although the lithographically micropatterned defects are larger than the natural surface defects, which are on the nanometer scale, we believe that understanding the behavior of these smooth simple defects could lead to a better understanding of contact angle hysteresis inherent to natural surfaces due to the *scale-independence* of the solution. Finally, the idea proposed here could be extended to periodic surfaces, e.g. hydrophobic surfaces, or even more complicated surfaces where, e.g. the function $h(x)$ is not symmetrical around $x = 0$, thus leading to complex wetting and dewetting of surfaces, i.e. a tunable contact angle hysteresis tailored to ones needs.

*steve.arscott@iemn.univ-lille1.fr

| | Experimental receding and advancing contact angles | | | | Model | |
|---|---|---|---|---|---|---|
| **Liquid** | $\theta_r$ | $\theta_a$ | $\theta_r^*$ | $\theta_a^*$ | $\theta_r^*$ | $\theta_a^*$ |
| H$_2$O [a] | 62.2°±2.8° | 91.8°±0.5° | 5.6°±1.3° | 147°±2° | 6.8 | 154.7 |
| H$_2$O [b] | 97.9°±2.3° | 114.1°±2.1° | 34.6°±3.4° | 159.3°±7.7° | 35.0 | 171.6 |
| C$_2$H$_6$O$_2$ [a] | 29.9°±1.5° | 55°±0.6° | 4°±0.7° | 119.2°±1.7° | 0.8 | 118.0 |
| C$_2$H$_6$O$_2$ [b] | 70.2°±1.3° | 123.9°±1.5° | 14.2°±1° | 152.6°±4.7° | 10.7 | 175.3 |
| IPA/H$_2$O (2.5/97.5) [b] | 88.5°±4.3° | 102.7°±1.4° | 24.3°±3.3° | 151.3°±4.5° | 25.5 | 164.5 |
| IPA/H$_2$O (5/95) [b] | 84.3°±1.8° | 98.3°±1° | 21.6°±3.2° | 147.7°±1.8° | 21.3 | 161.0 |
| IPA/H$_2$O (10/90) [a] | 44.5°±1.2° | 72.9°±1.1° | 4.9°±0.1° | 137.6°±4.6° | 2.3 | 135.9 |

Table. Experimental values for receding contact angle ($\theta_r$) and advancing contact angle ($\theta_a$) for liquids on SU-8[a] and SU-8 coated with a fluorocarbon (FC) layer[b]. The values were obtained using tilt and syringe methods using a DSA100 contact angle meter (Kruss, Germany). Table also shows Experimental values for the effective receding contact angle ($\theta_r^*$) and effective advancing contact angle ($\theta_a^*$) for various liquids on SU-8 and SU-8 coated with a fluorocarbon (FC) surfaces containing defects. The values were obtained using syringe and evaporation methods. Also shown are values predicted by the model (using $\kappa = 0.625$).